\documentclass[msom,nonblindrev]{informs3}

\DoubleSpacedXI

\usepackage{natbib}
 \bibpunct[, ]{(}{)}{,}{a}{}{,}%
 \def\bibfont{\small}%
\usepackage{subcaption}

\usepackage[colorlinks,
linkcolor=red,
anchorcolor=blue,
citecolor=blue
]{hyperref}

\TheoremsNumberedThrough     

\EquationsNumberedThrough    

\begin{document}




\TITLE{Cloud Electricity Storage for Multi-Service Battery Operation (Extended Version)}

\ARTICLEAUTHORS{%
\AUTHOR{Mohammad~Rasouli}
\AFF{Department of Management Science and Engineering, Stanford University, CA, 94305, USA, \EMAIL{rasoulim@stanford.edu}} 
\AUTHOR{Tao Sun}
\AFF{Department of Civil and  Environmental  Engineering, Stanford University, CA, 94305, USA, \EMAIL{luke18@stanford.edu}}
\AUTHOR{Camille~Pache, Patrick~Panciatici, Jean~Maeght}
\AFF{French transmission system operator \emph{Réseau de Transport d’Electricité} (RTE), 7C Place du Dôme, 92800, Puteaux, France, \EMAIL{camille.pache, patrick.panciatici, jean.maeght@rte-france.com}}
\AUTHOR{Ramesh Johari}
\AFF{Department of Management Science and Engineering, Stanford University, CA, 94305, USA, \EMAIL{rjohari@stanford.edu}}
\AUTHOR{Ram~Rajagopal}
\AFF{Department of Civil and  Environmental  Engineering, Stanford University, CA, 94305, USA, \EMAIL{ram.rajagopal@stanford.edu}}
} 

\ABSTRACT{%
 \textbf{Problem definition:} We study a \emph{cloud storage operator} who provides shared storage service for electricity end-users using the residual part of a multi-service grid-scale battery primarily used for high priority grid services. We design an optimal product offering, pricing and customer portfolio.
 \textbf{Academic/practical relevance:} A framework and solution approach for assessing and operating such multi-service battery operations with stochastic services and different priority levels is an open problem \citep{parker2019electric}. 
 \textbf{Methodology:} We model the problem as a two-stage stochastic optimization between high priority stochastic grid services and low priority cloud storage for stochastic end users. We also propose the operational metrics of \emph{multiplexing gain} and \emph{probability of blocking} to assess the operation of multi-service multi-user battery. 
 To address the computational challenge of solving the stochastic optimization with a large number of end-users, we propose \emph{effective capacity} as a convex approximation that allows an analytical solution. We then provide an empirical analysis based on real grid congestion data from RTE France, and a large dataset of end-users' electricity consumption in California. 
   \textbf{Results:}
 Our empirical analysis shows (i) our proposed effective capacity is a close approximation, (ii) battery operation and profit are sensitive to the cost of external resources, number of end-users, and  RTE's leasing price of the battery, and (iii) with only a slight discount of the leasing price ($\sim$ 1\%), the profit of the third party from a stochastic residual battery can be the same as that of a  deterministic one. 
 \textbf{Managerial implications: } Cloud storage as a low priority service can profitably exist alongside other high priority battery services, making integration of more storage in the grid economically viable, and allowing larger intermittent renewables, a key path towards reduced carbon emissions.}


\HISTORY{}

\maketitle
%


\section{Introduction} \label{sec:intro}

The rapid penetration of non-dispatchable, highly stochastic renewable energy sources like wind and solar into the traditional electricity grid requires significant development of flexible resources in the system, including energy storage. Recent technological developments, especially on electrical batteries, promise to enable a range of different services provided by energy storage in electricity systems (e.g., electricity arbitrage, frequency control, congestion management, etc.).  Given the highly stochastic nature of demands as well, a battery is most efficiently utilized if  its capacity is shared among different users for different services, while taking into account the potential conflicts between them in access to the battery. (See \cite{parker2019electric} and \cite{fitzgerald2015economics} for surveys on multi-service battery operation.) This is referred to as \emph{multi-service} use of the battery where the battery capacity is shared between users for multiple services.  \cite{parker2019electric} identifies the valuation of multi-service energy storage as one of the most fundamental open problems in the adoption of batteries in the grid; although it has been heavily studied in recent years, a defined architecture and set of metrics to plan, schedule, and evaluate the battery in a multi-service framework is missing.  Such architecture should also be evaluated empirically.

\begin{figure}[htbp]
\vspace{0.00mm}
\setlength{\belowcaptionskip}{-13pt}
\begin{center}
\includegraphics[width=9cm]{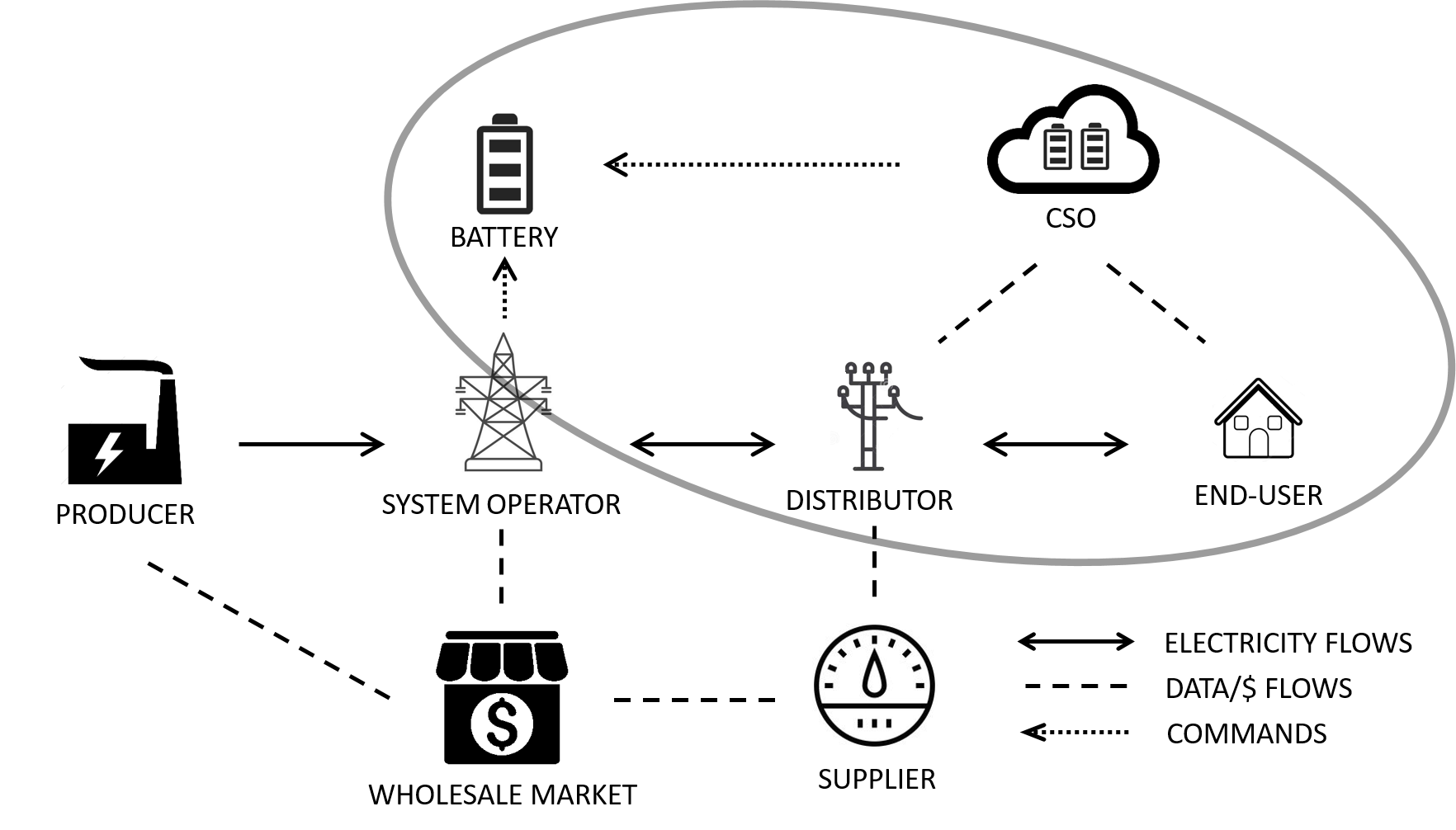}
\caption{Interactions between the different agents of the electricity system where the battery is used for cloud storage and grid services.}
\label{model}
\end{center}
\end{figure}

This paper addresses multi-service battery operation, specifically for the case of sharing battery between two services: grid congestion management and {\em cloud energy storage}.
Cloud storage is a storage service that allows different end users to have access to \emph{virtual storage} capacities contracted with a service provider, called a \emph{cloud storage operator} (CSO) who operates centralized large-scale batteries. 
Cloud storage has gained attention for the flexibility it provides for end-users; while end users operate these virtual batteries the same way they would do with their own individual battery, mainly for behind-the-meter energy arbitrage, they enjoy lower costs thanks to sharing and economy of scale as well as the flexibility convenience of cloud services (e.g., they can adjust their virtual storage capacities over time). Moreover, cloud storage does not have the problems of behind the meter batteries such as risk of fire, need for available space, or moving the battery at each relocation. On the other hand, cloud storage as a low priority service is suitable for a multi-service battery operation with high priority reliability services like grid congestion, frequency, or voltage control (see Section \ref{sec: RTE} for more discussion on technology and regulations for cloud storage). The agents and their interactions under cloud electricity storage multi-service setup are illustrated in Figure \ref{model}.

Two stochastic effects impact the potential efficiency gains of cloud storage. First, statistical multiplexing from the aggregation of diverse stochastic users in the battery allows for overbooking, which means the CSO can offer a virtual storage capacity greater than the one physically installed. Second, because of overbooking, there is the possibility that customers may be {\em blocked} when they want to access the battery; the CSO should have access to external resources to cover its commitment whenever the physical battery is blocked.  To capture these effects, we use \emph{multiplexing gain} and \emph{probability of blocking} as key metrics to assess the value of cloud storage.

We present our framework by studying large-scale batteries shared between two different services: high priority grid congestion management for the system operator, and low priority cloud storage for users. This is the use case of Réseau de Transport d'Électricité (RTE), the French Transmission System Operator (TSO), which will install three batteries of $24$MWh each by $2020$ \citep{cre2017deliberation}. RTE would like to keep the ownership of the battery for congestion management, while allowing resale of the residual share of the battery by third party profit-maximizing firms. Cloud storage is a reasonable model for third party provision of this residual capacity (see Section \ref{sec: RTE} for more discussion). 

We have two main contributions.  First, we formulate an optimization that determines the optimal planning and operation of cloud storage to maximize the CSO's profit while guaranteeing  high-reliability for the congestion management service. 
In principle, this can be a computationally challenging problem. 
 This computational complexity is due to the stochastic nature of the demand which results in non-convexity of penalty for \emph{blocking events} which are part of CSO's profit calculation.
We approximate the penalty of blocking events with \emph{effective capacity} by extending the concept of \emph{effective bandwidth} of a shared communication channel in the telecommunication literature. Effective capacity allows a convex approximation of the CSO's profit. Furthermore, while effective bandwidth is used for the communication bandwidth of each source, effective capacity here refers to the power capacity of the centralized battery. Furthermore the effective capacity allows for external resources to cover the resource overload, and the distribution of the users' battery schedule are time-variant. 

Our second contribution presents a calibrated empirical analysis of cloud storage with grid congestion management, using a large set of user consumption data in California and a real grid use case from RTE. We use unsupervised machine learning to cluster the users and estimate the distribution of the battery schedule per cluster. Our simulation confirms that maximizing the approximated CSO's profit using effective capacity provides a close solution to that of maximizing original CSO's profit. Furthermore, our findings show that cloud storage is an economically viable business for third parties, with two vital factors determining its profitability: (a) high levels of heterogeneity in end-users (for example with different net load profiles or electricity rates) to increase statistical multiplexing; and (b) affordability of external resources for covering blocking incidents. Finally, we provide a sensitivity analysis of the battery operation and profit to the cost of external resources, the number of end-users, and the RTE leasing price of the battery to the CSO. Interestingly we observe that with only a slight discount by RTE, the profit of the CSO from the stochastic residual battery (with congestion management) can be comparable to that of a deterministic battery (without congestion management).

\textbf{Organization:} The remainder of the paper is organized as follows. A literature review is presented in Section \ref{sec:literature}. Section \ref{sec: RTE} provides discussion on the regulations and technologies for cloud storage and explains the use case of RTE. Section \ref{sec:model} presents the model and the optimization problems. Section \ref{sec:metrics} defines the operational metrics of multiplexing gain and probability of blocking, and also the effective bandwidth. Section \ref{sec:empirical} provides a calibrated empirical analysis for cloud storage with congestion management. Finally, Section \ref{sec:conclusion} concludes and gives directions for future work.

\section{Literature review} \label{sec:literature}
{Single-service operation of battery is studied extensively (see \cite{parker2019electric} and \cite{fitzgerald2015economics} for a survey in operational research and engineering correspondingly). Some examples are arbitrage in retail or wholesale \citep{figueiredo2006economics,  sioshansi2009estimating, cruise2019control, lohndorf2013optimizing, sioshansi2011comparative, zhou2016electricity, paine2014market, jiang2015approximate, jiang2015optimal}, facilitating renewable energy integration \citep{harsha2014optimal, kim2011optimal, qi2015joint,ru2012storage, babrowski2016electricity,billionnet2016robust, greenblatt2007baseload, salas2018benchmarking}, home energy management \citep{vehvilainen2003managing, lin2020revisiting}, congestion management \citep{vargas2014wind}, virtual power plant (VPP) \citep{pudjianto2007virtual}, and cloud electricity storage \citep{liu2017cloud, zhao2019virtual}}. In this paper, we complement the use cases presented in \cite{liu2017cloud} and \cite{zhao2019virtual} by mixing in a multi-service with congestion management.

Sharing a single-service battery across multiple users is studied for behind-the-meter batteries in \cite{patel2018peer} and \cite{wu2016sharing}. {Optimal sharing of a single asset among multiple stochastic users is computationally challenging. Current studies of cloud storage in \cite{liu2017cloud} and \cite{zhao2019virtual} are limited to very specific assumptions. Similar sharing problem is studied in telecommunication to determine the communication capacity of a shared channel that keeps the probability of packet dropping below a threshold  by using \textit{effective bandwidth} approximations for the statistical multiplexing of various sources of communication signals (see   \cite{schwartz1996broadband} for a review). \cite{hui1988resource, hui2012switching} introduced this concept for a simple model of an unbuffered resource. The effective bandwidth concept was then generalized to a buffered resource with multi-class queues in \cite{kelly1991effective}. We borrow the ideas of effective bandwidth and extend them for sharing electricity batteries by considering time dependency of the battery schedules and constraints for both energy and power capacity. This way we introduce effective capacity. Similar approach in extending  existing results in telecommunication to energy scheduling problems is used in  \cite{jiang2016envelope} in the context of quality of service and renewable energy.}

Multi-service operation of battery has been studied in specific setups. At the transmission level, \cite{walawalkar2007economics}, \cite{drury2011value}, \cite{anderson2016cooptimizing}, \cite{wu2015energy}, and \cite{cheng2016cooptimizing} study the economic value and optimization of energy storage for two or more services to improve the efficiency and reliability of the grid including energy arbitrage in wholesale markets and ancillary services (such as regulation and reserve). At the distribution level, \cite{xi2014stochastic} and \cite{moreno2015milp} study the optimal schedule and control of distributed energy storage for distribution network congestion management in multi-service with retail energy price arbitrage, capacity, reserve, and frequency regulation. To the best of our knowledge, the idea of stacking multiple storage services with multiple users of each service has not been studied in the literature. We study this setup for the two services of congestion management and cloud storage.

\section{Cloud Electricity Storage}
\label{sec: RTE}
In this section we present the business case of cloud electricity storage based on new trends in the electricity grid.
First, behind-the-meter batteries for end-users are trending mainly to reduce their electricity bill by arbitrage against time-of-use retail prices; rooftop solar further increases such arbitrage opportunities especially in places where the retailer does not compensate the rooftop solar generation injected to the grid at high prices. Reducing the solar compensation, however, is a trend not only in France \citep{france_solar}, but also California (NEM 2.0) \citep{gong2017financial} and New York (Value Stack program) \citep{astoria2017radicality} because high compensation is hard to sustain for the retailer. 

Another complementary trend is cloud storage business where the CSO
contracts different end-users to have access to \emph{virtual storage} capacities instead of installing their physical battery in house. Cloud storage has a higher value compared to distributed behind-the-meter batteries when the new costs it imposes, which are distribution costs and line-losses, do not exceed the cost reduction due to battery economy of scale and sharing multiplexing gain.
Extra costs in the form of distribution costs and line losses are both because the battery is not located at the same place as the end-users. Therefore, in contrast to behind-the-meter batteries, charging the virtual battery with excess rooftop solar is subject to using distribution lines; this adds both service costs by the distributor and line losses. However, both of these costs are typically small and are ignored in this paper. Note that further if there are no locational marginal prices (e.g. the case of France empirically studied in this paper), battery charging at two different locations at the same time will incur no price difference for a market participant. In addition, the implementation of cloud storage requires virtual metering, which is a billing scheme such that the operation of virtual batteries in remote can be counted as users operating their own behind-the-meter batteries in their bill. Similar virtual metering systems are provided in Australia, California, and Greece to facilitate solar systems shared across multiple users \citep{langham2013virtual, poullikkas2013review, mavromatakis2018photovoltaic}. Moreover, although not considered in this paper, cloud storage has the added value of flexibility and convenience that does not exist in behind-the-meter batteries.

An important problem in Cloud Electricity storage is that since the sum of the virtual batteries can exceed the physical battery, there are occasions where users require access to their virtual batteries simultaneously beyond the level of the physical battery. Hence, the CSO should have access to external resources that can cover those incidents. The CSO can not use the wholesale market as an external resource for covering service for end-users (this way the CSO can arbitrage wholesale prices against retail prices which is not allowed as mentioned earlier), but the CSO can access the retailer as an external resource leveraging the virtual metering. 

Finally, Cloud Electricity Storage is apt for the multi-service with other high priority services as we discuss in detail for the case of RTE in Section \ref{ssec: case of rte}. Access to external resources becomes also important for using cloud storage in multi-service; in using stochastic residual battery by CSO, that there are occasions the battery is taken by the other high priority services (e.g. congestion management) and therefore CSO can not deliver service to end-users. These occasions should be covered by the external resources.

\section{Model and Optimization Problems} \label{sec:model}
We model the multi-service battery operation of high priority service and low priority Cloud Electricity Storage. For the high priority service we consider the grid congestion management which is the case of RTE (the same model can be used for other types of stochastic high priority services). Grid congestion management is a stochastic single user service operated by the system operator that overrides any other user of the battery. Furthermore, the residual share of the battery is operated by the CSO for cloud storage business. We model the cloud storage business as a stochastic multiple-user service and as a two-stage game between the CSO and the users {where in the first stage the platform offers a menu of contracts in the form of a size of the virtual battery and a subscription fee, and in the second stage each user chooses its best contract from the menu. Then over the horizon of $1$ to $T$, users choose how to charge/discharge their virtual battery and CSO chooses how to operate the central battery}. We model this two-stage game with a two-stage optimization. In Section \ref{ssec: agents} we provide a model. Then, we present the CSO's and the users' optimization problems in Section \ref{ssec: problems}.

\subsection{Model}\label{ssec: agents}

Consider time horizon $T$ (each time step is one hour). Consider a system of a wholesale market, an electricity supplier (retailer), a system operator, a distributer, a cloud storage operator (CSO), and $N$ end-users e.g. households (Figure \ref{model}). The retailer buys electricity in the wholesale market and sells at retail price to end-users through the distribution network. We assume the distribution service costs are reflected on the retail price to end-users. The system operator controls the congestion management, among other reliability responsibilities, and sends proper commands to the physical battery. The CSO has access to the residual share of the battery and provides virtual battery service to end-users through the distributor. The CSO also has access to external resources (e.g. retailer or locally owned generators) for purchasing electricity; however, we assume the CSO is not compensated for selling back electricity to the retailer.
 
 The virtual battery service is provided by the CSO to users (households) and each user decides how to operate his virtual battery. We use superscript $h$ to refer to user-related variables. The load of user $i$ is inelastic and is denoted by $d^h_{i,t}$ at hour $t$. The user's distributed generation (rooftop solar) at time $t$ is $g^h_{i,t}$. $d^h_{i,t}$ and $g^h_{i,t}$, $t \in T$ form a random process.  $(d^h_{i,t} - g^h_{i,t})$ is the net demand of  user $i$ at time $t$ without considering the virtual battery operation. We denote the energy capacity of the virtual battery of user $i$ by $c^h_i$ and its power capacity by $r^h_i$. At time $t$, the virtual battery operation is defined by a virtual state of charge $s^h_{i,t}$ and the power charged/discharged in the virtual battery is denoted by $a^h_{i,t}$. We denote the space of $a^h_{i,t}$ by $A^h_{i,t}$. We assume users can use their virtual battery for energy arbitrage beyond their net demand, meaning they can store electricity in their virtual battery at cheap retail rates to use during periods with higher retail rates. Thus, the difference $(d^h_{i,t} - g^h_{i,t} + a^h_{i,t})$, denoted by $\tilde{d}^h_{i,t}$, is the amount that user $i$ is billed by the electricity supplier at time $t$, at a price denoted by $p^h_t$. We assume $p^h_t$ is equal to the retail price (e.g. time-of-use (ToU) tariff). In this way, the virtual battery is similar to a behind-the-meter battery from a user's perspective (For the sake of simplicity, we exclude fixed charges in the tariff when computing users' electricity bills.). We allow the retail price to be different for buy and sell of electricity from/to the grid which we denote by $p^{r+}_t$ for buying from grid and $p^{r-}_t$ for selling to grid. User $i$'s total payment over period $T$ is denoted by $\kappa^h_i$ and is equal to
\begin{align}
\kappa^h_{i}= \sum_{t \in T} \bigg ( & p^{r+}_t \times [\tilde{d}^h_{i,t}]^+ - p^{r-}_t \times [\tilde{d}^h_{i,t}]^- \bigg )
\end{align}
where $[.]^+$ is the positive part and $[.]^-$ is the negative part so that, for any $z \in \mathbb{R}$, $z=[z]^+ - [z]^-$. 
 
We can now model user's strategy for operating the battery. The information history of user $i$ by time $t$ is denoted by 
\begin{align}
    x^h_{i,t} =(s^h_{i,0}, a^h_{i,1}, ...,a^h_{i,t-1},d^h_{i,1},...,d^h_{i,t},g^h_{i,1},...,g^h_{i,t}).
\end{align}
Consider the space of $x^h_{i,t}$ to be $X^h_{i,t}$. Then, the user's strategy of running the virtual battery denoted by $\sigma^h_{i,t}$ is a function that maps $X^h_{i,t}$ to $A^h_{i,t}$. We denote by $\Sigma^h_{i,t}$ the set of all such strategies.

We denote the  energy capacity of the battery owned by the CSO by $c^b$ and its power capacity by $r^b$ (superscript $b$ is for referring to CSO-related variables as opposed superscript $h$ for household-related vairables). We denote the CSO's annual cost of battery by $C^b_{eq}(c^b,r^b)$ (this includes investment cost in the form of renting paid to RTE, operation, and maintenance) of a battery of energy capacity $c^b$ and  power capacity $r^b$ calculated over the battery lifetime (We assume the operation and maintenance costs are fixed costs). We use a linear form for $C^b_{eq}$. This linear cost is supported practically as discussed in Section \ref{sec:data}. We also assume  large-scale batteries benefit from economy of scale, that is, the CSO's cost of battery is lower than the cost of distributed batteries with the same aggregate size. In the case where the battery is used primarily for congestion management, the system operator sends signals to the battery to allocate sufficient capacity and power resources for congestion management. Therefore, the residual share of battery capacity available for cloud storage varies with time which we denote by $c^b_t$ and $r^b_t$. Note that the residual share of battery capacity is usually informed by the system operator some time ahead of the congestion happening. Thus, it is known to CSO at the time of operation. However, it is stochastic to CSO at the time of planning. Note that we assume at the occasions of congestion management, the CSO has still the feasibility to provide electricity to users; in other words the grid congestion management does not impact the delivery service to users (this could be that the congested lines do not impact users lines to battery, or CSO has access to external local resources not impacted by the grid congestion).

At time $t$, the battery operation is defined by a state of charge $s^b_t$ and the amount of electricity charged or discharged $a^b_t$. We assume that the battery round-trip efficiency is close to $1$ (Round-trip losses may be included for a more detailed empirical analysis, but they do not change our major results.). The CSO has access to external resources with prices $p^b_t$ We assume the CSO is charged for buying electricity from external resources but does not get paid for selling electricity back (There is an on going regulatory discussion about the operation of grid batteries. Our model and study can be easily extended to allow the CSO to sell back electricity. Note also that the retailer is one external resource the CSO can have access to, alternatively the CSO can have local generators.). The CSO's total payment for external resources over period $T$ is denoted by $\kappa^b$ and is equal to 
\begin{equation}\label{eq: kappa b}
\kappa^b= \sum_{t \in T}p^b_t\times  [a^b_t -\sum_{i\in N} a^{h*}_{i,t} ]^+
\end{equation}


Now we can model the CSO's strategy. It involves an initial decision on the size of the battery and contracts, followed by operation of the battery in the entire horizon. At time $t$, the CSO's information history is $x^b_t=(s^b_{0},a^b_1,...,a^b_{t-1}, \{s^h_{i,0}, a^h_{i,1}, ...,a^h_{i,t}\}, \forall i\in N, \{c_t^b,r_t^b\}, \forall t\in T)$. This information history includes all previous commands received from the users as well as the previous operations of the centralized battery.  We denote the space of $x^b_t$ with $X^b_t$ and the space of $a^b_t$ with $A^b_t$. Then, the CSO's strategy of running the battery $\sigma^b_t\in \Sigma^b_t$ is a function that maps $X^b_t$ to $A^b_t$, where $\Sigma^b_t$ is the set of such strategies.
\begin{align}\label{eq: cso strategy}
    \sigma_t^b: X_t^b\rightarrow A_t^b.
\end{align}

Note that in the above model, as mentioned in Section \ref{sec:intro}, we assume the distribution and transmission fees in the user's bill are negligible. We also assume the virtual metering is available for facilitating the cloud storage business.

We are now ready to present the two-stage game between CSO and users.

\subsection{Optimization Problems}\label{ssec: problems}
We model the CSO and users interaction by a principal-agent model in the form of a two-stage game. The CSO first decides the size of the battery he wants to install, and what menu of virtual battery contracts to propose to the users. Each virtual battery contract in the contract menu $Q = \{(q_k, c_k, r_k)\}_{k\in K}$ is in the form of energy and power capacity $(c_k, r_k)$ and price $q_k$ for access in the entire horizon $T$.  Each user then decides how much virtual battery to rent and how to use it. Then, the CSO schedules the battery operation considering the signed contracts and users' use of their virtual batteries.

We first study the users' problem. User $i$ chooses its virtual battery size $(c^h_i, r^h_i)$ from the contract menu $Q = \{(q_k, c_k, r_k)\}_{k\in K}$ and its optimal strategy for operating the virtual battery ${\sigma^h_{i,t}}^*, t\in T$. In this decision the user has to find a trade-off between the savings it gains from energy arbitrage on its electricity bill through the use of its virtual battery and the cloud storage fee. Optimization (\ref{eqn:Opt_user}) formulates users decision. The first part of the objective function (\ref{Opt_user1}) corresponds to the cost associated with the contracted virtual storage capacity, while the second part is the expected electricity bill. Here the expectation is taken over external random states for the inelastic demand and for rooftop solar production. The expected electricity bill decreases when the storage capacity increases.
\begin{subequations}
\label{eqn:Opt_user}
\begin{align}
\pi_i^{h*}= & \min_{(q^h_i,c^h_i,r^h_i) \in Q, \sigma^h_{i,t}\in \Sigma^h_{i,t}} q^h_i + E [\kappa^h_i]\label{Opt_user1}\\
\mbox{subject to} \;  & -r^h_i\leq a^h_{i,t}\leq r^h_i \label{Opt_user2}\\
& 0\leq s^h_{i,t}\leq c^h_i \label{Opt_user3}\\
& s^h_{i,t} = s^h_{i,t-1}+a^h_{i,t} \label{Opt_user5}
\end{align}
\end{subequations}
The first constraint (\ref{Opt_user2}) corresponds to the maximum charge and discharge rates, while the second constraint (\ref{Opt_user3}) relates to the minimum and maximum states of charge. Finally, (\ref{Opt_user5}) defines the dynamics of the battery. The solution to the above problem gives $c^{h*}_i, r^{h*}_i, \sigma^{h*}_{i,t}$ which determine the virtual battery characteristics and its use by the user.

We next present CSO's problem which determines the contract menu $Q$ and the optimal battery investment and operation. The CSO may offer more virtual storage capacity than physically available in order to take advantage of the multiplexing gain between different users. However, there can be times where there is a mismatch between the aggregated virtual command $\sum_{i}a^h_{i,t}$ and the battery command $a^b_t$. In these cases, the CSO needs to cover the mismatch from external resources at price $p^b_t$.  Thus, the CSO's problem is given by the following optimization problem. This optimization trades off between the battery investment cost and the mismatch penalty from external resources.

\begin{subequations}
\label{eqn:Opt_cso}
\begin{align}
\pi^{cso*}= & \max_{c^b \in \mathbb{R}^{+}, r^b \in \mathbb{R}^{+}, (q^h_i,c^h_i,r^h_i) \in Q, \sigma^b_t \in \Sigma^b_t, Q}   -C^b_{eq}(c^b,r^b)
 + \sum_{i\in N}q^h_i - E[\kappa^b] \label{Opt_cso1}\\
\mbox{subject to} \;  & -r^b\leq a^b_t \leq r^b\label{eq:power_cap_cons}\\
& 0\leq s^b_t\leq c^b\label{eq:energy_cap_cons}\\
& s^b_t = s^b_{t-1}+a^b_t
\end{align}
\end{subequations}
The first part of the objective function (\ref{Opt_cso1}) corresponds to the equivalent battery investment cost over the considered period, the second part is the revenue from the contracts $i\in N$, then the last part is the expected cost from the mismatches between the virtual command and the battery schedule. Here the expectation is taken over the randomness of battery schedule of each user. For instance, if the CSO has access to external resources at price $p^b_t$, when the battery is empty and users want to discharge their virtual storage, the CSO will buy the missing energy at price $p^b_t$.  In the case where the CSO does not have access to external resources, it should ensure that the battery capacity is large enough to follow the aggregated virtual command at all times. 

Formulation of Equation \eqref{eqn:Opt_cso} is with a single use case of the battery, only cloud storage. To extend the problem to incorporate high priority grid congestion management we do the following steps. First we replace  constraints (\ref{eq:power_cap_cons}) and (\ref{eq:energy_cap_cons}) with
\begin{align}\label{eq: residual power}
r^{b,l}_t\leq a^b_t \leq r^{b,h}_t\\
 0\leq s^b_t\leq c^b_t.\label{eq: residual energy}
\end{align}
where $r^{b,l}_t$ and $ r^{b,h}_t$ are the lower and upper bounds for $a^b_t $ respectively.   This formulation sets aside capacity for congestion management and allocates the residual share to cloud storage. In this way, it gives high priority to congestion management.  $r^{b,l}_t$, $ r^{b,h}_t$, and $c^b_t$ are considered stochastic at the time of CSO's investment. Note that, meeting stochastic constraints of (\ref{eq: residual power}) and (\ref{eq: residual energy}) requires CSO's strategy for setting $a_t^b$ to be adaptive to the realizations of $r_t^{b,h}, r_t^{b,l}$ and $c_t^b$. However, solving numerically for the optimal adaptive policy is complicated. Therefore, we propose an alternative problem by considering CSO's strategies introduced in (\ref{eq: cso strategy})  which is not adapted to uncertainty of $r_t^{b,h}, r_t^{b,l}$ and $c_t^b$, and by formulating (\ref{eqn:Opt_cso}) a \textit{chance constrained optimization} where  constraints (\ref{eq: residual power}) and (\ref{eq: residual energy}) are replaced by 
\begin{subequations}
\label{eq: chance_cons}
\begin{align}
&\textbf{Prob}(r^{b,l}_t\leq a^b_t)\geq \eta \label{eq: chance_cons_a1}\\
&\textbf{Prob}(a^b_t \leq r^{b,h}_t)\geq \eta\\
&\textbf{Prob}(s^b_t\leq c^b_t)\geq \eta\\
& 0 \leq s^b_t
\end{align}
\end{subequations}
Here $\eta$ is the confidence level of a meeting a constraint. The chance constraints in (\ref{eq: chance_cons}) are linear. To see this, consider that (\ref{eq: chance_cons_a1}) is equivalent to
\begin{equation}
\label{eq:emp_quantile}
F^{-1
}_{r_t^{b,l}}(\eta) \leq a^b_t
\end{equation}
where $F_{r_t^{b,l}}$ is the cumulative distribution function of $r_t^{b,l}$, and similarly for the other constraints in (\ref{eq: chance_cons}).

\section{Solution approach and operational metrics} \label{sec:metrics}
Solving the two-level optimization problems (\ref{eqn:Opt_user}) and (\ref{eqn:Opt_cso}) in Section \ref{ssec: problems} is computationally challenging since the objective function of (\ref{eqn:Opt_cso}) does not have a closed concave  analytical form. Moreover, a solution to the two-level optimization problems (\ref{eqn:Opt_user}) and (\ref{eqn:Opt_cso}) does not provide useful intuitions for operation of the multi-service battery. We address these two issues in this section.

The first challenge is because the expected blocking costs $E[\kappa^b]$ does not have a closed convex analytical form considering \eqref{eq: kappa b}. This in turn is due to the stochastic nature of users' battery schedules, and the fact that the penalty is a function of distribution tail. To address the issue, we offer the method of effective capacity which provides an closed-form convex approximate to the expected blocking cost $E[\kappa^b]$ in  (\ref{eqn:Opt_cso}) using.

The solution to the two-level optimization problem between the CSO and users formulated in (\ref{eqn:Opt_user}) and (\ref{eqn:Opt_cso}) does not provide a general intuition of the operation of the battery beyond the bottom line profit value. In other words, the operation of the battery is encoded through the large space of strategies for the CSO and end-users. Therefore,
we define the multiplexing gain and the probability of blocking as intuitive and simple operational metrics for understanding the value of a battery in a general multi-service framework. Moreover, these metrics are also correlated with the CSO's profit: the higher multiplexing gain means the higher the profit, and blocking probability is  sensitive to the CSO's penalty costs from external resources. We use these metrics for our empirical study of cloud storage to shed light on the operation of the battery.

\subsection{Effective Capacity}
\label{sec:het_users}

{The CSO's problem (\ref{eqn:Opt_cso}) proposed  in Section \ref{ssec: problems} is a  stochastic sequential optimization. In this section, we  focus on a convex approximation of the problem which allows for fast solution in contrast to the computationally expensive Monte Carlo method; to this end, we propose the concept of \textit{effective capacity} by providing a closed form convex approximation to the expected blocking cost $E[\kappa^b]$ in \eqref{eq: kappa b} as a function of the physical battery schedules $a^b_t$.} Note that $E[\kappa^b]$ is convex but does not have a closed form considering the stochasticity of end-user battery schedules. This makes the CSO's optimization problem challenging computationally.

We extend the model in Section \ref{ssec: problems}, by assuming there are $J$ classes of users with $n_j$ being the number of users in classes $j$. Each type of users has a different distribution for electricity consumption and thus different battery schedules from problem (\ref{eqn:Opt_user}). Denote $a_{i,t}^{h,j}$, $i=1,...,n_j$  to be the charging power of user $i$ in type $j$ at time $t$. Here we denote the total virtual battery (discharging) commands with 
\begin{align}
A_t=-\sum_{j=1}^J \sum_{i=1}^{n_j} a_{i,t}^{h,j}.
\end{align}
With large deviation method, when $n=\sum_j n_j$ becomes large we approximate $E[\kappa^b]$ by 
\begin{align}
\label{eq:bat_mismatch}
E[\kappa^b]\approx \sum_{t \in T}p^b_t\times  [a^b_t -\sum_{j=1}^J \sum_{i=1}^{n_j} \mu^{h,j}_{i,t} ]^+ + R_t(-a_t^b)
\end{align}
where 
\begin{align}
\mu^{j}_{t}=E[a_{i,t}^{h,j}],\\
R_t(-a_t^b)=E[A_t+a_t^b]\label{eq: Rtb}.
\end{align}
$R_t$ represents the expected amount of summed users' discharging that exceeds the discharging value of the central battery. Assuming $E[A_t]<-a_t^b$ and $P\{A_t>-a_t^b\}>0$, the Chernoff's bound gives
\begin{align}
\log P\{A_t\geq -a_t^b\}\leq \log E[e^{\tau(A_t+a_t^b)}]
\end{align}
To provide a analytical approximation for the $\log E[e^{\tau A_t}]$, we use the following assumption.

\begin{assumption}\label{ass: ind log moment} We assume $a_{i,t}^{h,j}$, $i=1,...,n_j$ are independent random variables across agents, with identical distribution across agents in a cluster $j$ at a certain time $t$.
\end{assumption}
Assumption \ref{ass: ind log moment} exists is used in the effective bandwidth literature \citep{kelly1995mathematical}. Note that practically for electricity batteries, $a_{i,t}^{h,j}$s are not independent across agents for example due to correlations through weather condition, prices, etc. This makes the problem difficult because bounding the sum of correlated random variables in an analytical form is challenging. From Assumption \ref{ass: ind log moment}, denoting the logarithmic moment generating functions by
\begin{equation}
M^j_{t}(\tau) = \log E[e^{-\tau a_{i,t}^{h,j}}].
\end{equation} 
we have 
\begin{align}
  \log  E[e^{\tau A_t} ]=\sum_{j=1}^J n_j M^j_{t}(\tau)
\end{align}
So intuitively the stronger the correlation across users battery operation, the lower is the accuracy of our approximation method. $\log P\{A_t\geq -a_t^b\}$ can be approximated by $\text{inf}_{\tau}[\sum_{j=1}^J n_j M^j_{t}(\tau)+\tau a_t^b]$, which will be an upper bound under Assumption \ref{ass: ind log moment}. Also, by large deviations approximation, when $n=\sum_j n_j$ becomes large, $\text{inf}_{\tau}[\sum_{j=1}^J n_j M^j_{t}(\tau)+\tau a_t^b]$ provides a tighter approximation of $\log P\{A_t\geq -a_t^b\}$. The effective bandwidth literature \citep{kelly1995mathematical} only approximates the probability of blocking (packet drop in communication) which is used to obtain an effective bandwidth for each source. In our case, however, we use further approximate $R_t$ which is the expected amount of blocked virtual battery schedule as follows. First, the CDF of $A_t$ can be approximated by
\begin{equation}
\label{eq: cdf}
F_t(x) \approx 1-\exp(\text{inf}_{\tau}[\sum_{j=1}^J n_j M^j_{t}(\tau)+\tau -x]).
\end{equation}

\begin{assumption}\label{ass: subexpo}
Assume the distribution of $-a_{i,t}^{h,j}$ is  sub-exponential  with parameters $(\nu^j_t, b^j_t)$. 
\end{assumption}

From Assumption \ref{ass: ind log moment} and \ref{ass: subexpo},  $A_t$ is a sum of independent sub-exponential random variables, and therefore following the CDF in (\ref{eq: cdf}) its PDF can be approximated as follows \citep{boucheron2013concentration},
\begin{equation}
f_t(x) \approx \frac{1}{2 b^{max}_t}\exp{(\frac{-\sum_j n_j\mu^j_{t}-x}{2 b^{max}_t})}\label{eq: approx At}
\end{equation}
where 
\begin{align}
    \mu^j_{t} &= E[a_{i,t}^{h,j}],\\
    b^{max}_t&=\max_j \{ b^j_t\}.
\end{align} 
Using \eqref{eq: approx At} above, $R_t(\underline{a}_t^b)$ in \eqref{eq: Rtb} can be approximated as 
\begin{align}\label{eq: approx Rt}
R_t(-a_t^b) \approx \int_{-a_t^b}^\infty (a+a_t^b) f_t(-a_t^b)da
=2b^{max}_t \exp (\frac{-\sum_j n_j\mu^j_{t}+a_t^b}{2b^{max}_t}),
\end{align}
which is a closed form analytical convex function of  $a_t^b$. Solving the CSO's problem in (\ref{eqn:Opt_cso}) thus becomes a convex optimization problem with a closed-from convex objective function.

\begin{definition}
{The \textbf{effective capacity}  for the cloud storage is the optimal energy and power capacity $(c^{b*}, r^{b*})$ of problem (\ref{eqn:Opt_cso}) when the expected blocking cost $E[\kappa^b]$ is approximated by (\ref{eq:bat_mismatch})} and $R_t(-a_t^b)$ is approximated with \eqref{eq: approx Rt}. 
\end{definition}

\subsection{Multiplexing gain and probability of blocking}
\label{ssec:oper_metrics}

It is important for the battery operator to have a measure of how the battery is used with simple metrics. One important feature is the overbooking of the battery that is the difference of real battery capacity and the aggregate virtual battery capacity. This particularly identifies the economy gain from sharing. On the other hand, virtual metering of a rush for using the shared resources there will be blocking, which is covered by external resources. Therefore, the other important feature is how much the battery operator relies on external resources for covering the costs. Hence we define operational metrics of multiplexing gain and probability of blocking.

We define the multiplexing gain of the battery as follows.
\begin{definition}
The \textbf{multiplexing gain} of a battery with storage capacity $c^b$ which provides service to $N$ different users with virtual capacity $c_i$ assigned to user $i$ is \begin{align}G=\frac{\sum_{i \in N} c_i-c^b}{\sum_{i \in N} c_i}.\end{align}
\end{definition}
The multiplexing gain is between zero and one with higher values meaning more economy of sharing. If there is no physical battery, the multiplexing gain is equal to one. If $c^b=\sum_{i\in N}c_i$ the multiplexing gain is zero and there is no economy of sharing.

Next, we define the probability of blocking as a measure of the battery unavailability to cover the virtual commands.
\begin{definition}
Consider a battery providing service to $N$ different users with virtual operation of user $i$ at time $t$ denoted by $a_{i,t}$, and the operation of the battery at time $t$ denoted by $a_t^b$. The  probability of blocking  at time $t$ is
\begin{align}
    P\bigg(|\sum_{i\in N} a_{i,t} - a^b_t| > 0\bigg),
\end{align}
and the probability of blocking is
\begin{align}
    \frac{1}{T}\sum_{t\in T} P\bigg(\bigg[|\sum_{i\in N} a_{i,t} - a^b_t| > 0\bigg]\bigg).
\end{align}
\end{definition}
Note that probability of blocking is a first order metric to measure blocking events; a complete description is the entire distribution of the blockings. Such distribution shows the severity of the blocking once they happen. 

\begin{figure}[htbp]
\vspace{-5pt}
\setlength{\belowcaptionskip}{-10pt}
\begin{center}
\includegraphics[width=8cm]{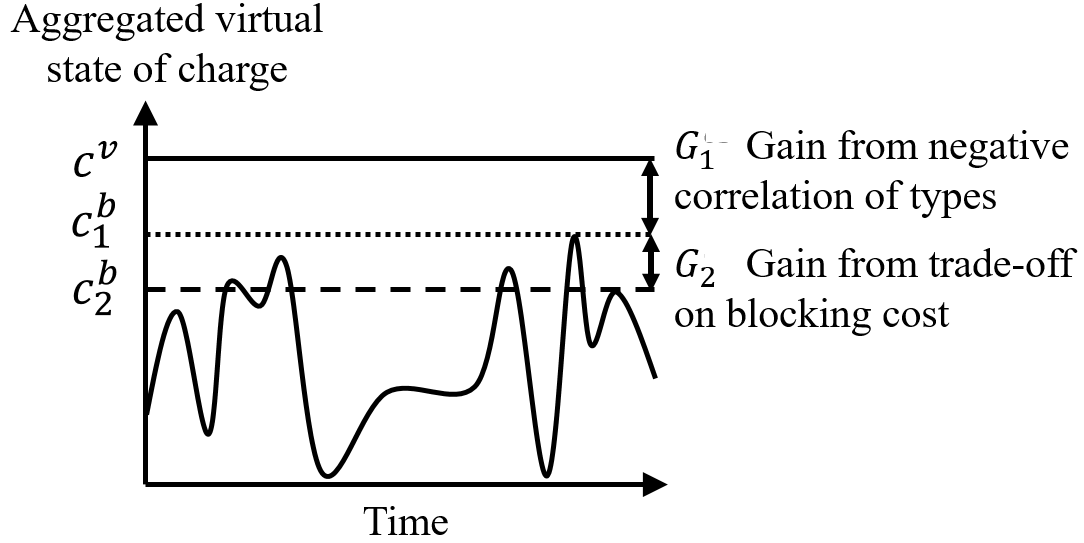}
\caption{Illustration of the multiplexing gain for cloud storage with and without access to external resources for covering blocking events.}
\label{mplx}
\end{center}
\end{figure}

Figure \ref{mplx} illustrates the multiplexing gain and its relation to the blocking probability in the case of cloud storage (the battery capacity is deterministic in this figure). This figure shows the aggregate virtual capacity of the battery $c^v=\sum_{i \in N} c_i$, and its state of the charge over time. When the CSO does not have access to external resources and consequently the probability of blocking is zero, the capacity of the battery should meet the peak of virtual state of charge, $c^b_1$. In this case the multiplexing only comes from the negative correlation between some users' operation of the battery ($G_1$). When  the CSO has access to external resources the CSO can trade-off between the cost related to the blocking events and the battery investment cost to decide the optimal capacity of the battery. This results in an increased multiplexing gain, ($G_2$); this multiplexing gain is a function of the users' statistical multiplexing as well as the cost of blocking (electricity price in external resources). 

Also, in the case where the CSO does not have access to external resources, the blocking events are caused by limiting constraints on the physical battery only. They correspond to the times where the battery power rate is limiting (either charging or discharging), when the battery is full and cannot be charged or when it is empty and cannot be discharged. The probability of blocking at time $t$ can is then
\begin{align}
P\bigg(& \bigg[\sum_i a_{i,t} > r^{b,h}_t\bigg]\cup \bigg[\sum_i a_{i,t} < r^{b,l}_t\bigg]\cup \bigg[s^b_{t-1}+ \sum_i a_{i,t} > c^b_t\bigg] \nonumber\\
& \cup \bigg[s^b_{t-1}+ \sum_i a_{i,t} < 0\bigg]\bigg)
\end{align}
However, if the CSO has access to external resources, the CSO may choose not to follow the aggregated virtual schedule which changes the probability of blocking.

\section{Empirical Study} \label{sec:empirical}

In this section we present and empirically study the case of RTE for multi-service battery for Cloud  Storage and grid congestion management. The details of the RTE case is presented in Section \ref{ssec: case of rte}. The data we use includes large-scale user demand and grid congestion management data described in Section \ref{sec:data}. We also cluster user demand data (Section \ref{sec: emp_cluster}) and use the obtained clusters to show the user optimization problem for contracting and operating individual batteries (Section \ref{ssec: user opt empiric}). 

Next, we first study the cloud storage as a single service both with  and without  CSO's access to external resources (Section \ref{ssec:with_without_ext}). We compare cloud storage business with alternative services (Section \ref{ssec:comp_other_cases}) and discuss the sensitivity of the results with respect to cost of external resources and number of users (Section \ref{ssec:sens_emp}). Next, we present the CSO's  problem in the multi-service setup of cloud storage and congestion management (Section \ref{ssec: cloud and congestion empirics}), and assess the sensitivity of results to the varying battery leasing costs to the CSO.

\subsection{Case of RTE}\label{ssec: case of rte}
We present the case of multi-service multi-user battery operation for  RTE. In the demonstrator project Ringo \citep{cre2017deliberation}, RTE planned to install a large capacity of batteries by 2020 to manage grid congestion in response to high renewable penetration (Ringo project phase 1). However, large scale batteries are new developments in the electricity grid and there is a regulatory discussion if RTE as a non-market participating entity is allowed to own a battery. Consequently, RTE is asked to pass the battery ownership to third parties (in 2023) who can use it to participate in the markets for providing ancillary and balancing services alongside congestion management (Ringo project phase 2) \citep{rte_projects}. Nevertheless,  with grid congestion occurring occasionally, 
RTE would like to pass the stochastic residual share of the battery (after fulfilling congestion management) to a third party for market-related services. RTE is not a profit seeker and will not lease the battery to the third party at a price larger than the original investment cost, but probably less since its availability is stochastic. The third party however is a profit maximizer (not looking to maximize social welfare or efficiency in the electricity system). Therefore, to make its case, RTE should prove a profitable business model exists for such third-party ownership of the residual battery. This is a challenging task, first because ancillary services are  high priority, and providing them with the residual battery will not be profitable due to high penalty costs. Second, the battery is not allowed to operate as a virtual retailer/aggregator arbitraging between the wholesale market and the existing retail prices. Moreover, arbitrage opportunity in the French wholesale market may not be high enough for covering the cost of the battery. Consequently, RTE would like to assess the profitability of using the residual battery to provide low priority behind-the-meter services for end-users in a sharing economy, known as \emph{cloud storage} business \citep{liu2017cloud} for a third party \emph{cloud storage operator (CSO)} (Figure \ref{model}).

\subsection{Data}\label{sec:data}

Our empirical study is based on real data from smart meter, solar irradiance, electricity price and tariff, and grid congestion.

The user electricity consumption data, which represents hourly consumption of $116170$ residential smart meters for a one-year period from August $2010$ to July $2011$, comes from customers of Pacific Gas and Electric Company (PG\&E) in Northern California. In addition to hourly electricity consumption data, the data set also contains information on the climate zone and zip code.

Historical solar irradiance data is available for the users at the zip-code level. Rooftop PV systems are sized according to  the California Public Utilities Commission (CPUC)’s zero net energy goals (The average annual electricity consumption in a building is equal to the yearly renewable generation on site \citep{cpuc2015netzero}).

PG\&E's ToU rate (We consider PG\&E’s E-TOU Option B in \cite{pge2016tou}. Alternative tariffs including tiered tariff system is not considered in this paper.) is used for retail electricity tariffs (see Table \ref{tou_tariff}). During holidays and weekends, all hours are charged the off-peak rate. If the CSO turns to retailer for external resources, his cost of external resources (that is, the penalty price) is at retail prices.

\begin{table}[htbp]
\caption{PG\&E's E-TOU Option B}
\centering
\begin{tabular}{c@{\hskip 0.1in}c@{\hskip 0.1in}c@{\hskip 0.1in}c}
\hline \hline 
Period & Off peak hours & Peak hours (4pm-9pm) \\ [0.5ex]
\hline
June-September & \$0.25511/kWh & \$0.35817/kWh \\  [0.5ex]
October-May & \$0.20191/kWh & \$0.22071/kWh \\  [0.5ex]
\hline
\end{tabular}
\label{tou_tariff}
\end{table}

We use the current battery investment cost function from a report by RTE  \citep{rte2017rei}, where $c^b$ is the energy  capacity and $r^b$ is the power capacity of the battery:
\begin{align} \label{eq: battery_cost}
& C^b(c^b,r^b)= 175[k\$/MW]\times r^b+395[k\$/MWh]\times c^b
\end{align}
Note that this cost function is only valid for large-scale batteries. Small individual batteries such as the Tesla Powerwall have a larger cost per kWh (Using the cost data from Tesla Powerwall, the total cost of installing $100$ Powerwall of $13.5$kWh each would range between $700$k\$ and $820$k\$, while installing a large battery of the equivalent size would cost $660$k\$.), and do not benefit from economy of scale.

We use  RTE's data to assess the residual capacity of the battery which will be available for cloud storage. In this study we had access to RTE simulation data where large batteries are used for grid congestion management. Grid congestion data is very specific to the studied zone, however some generic parameters causing congestion (e.g. wind or temperature) can be extracted and correlated to the cloud storage operation. RTE has selected three zones at the sub-transmission grid ($63$kV and $90$kV) for the installation of large batteries by $2020$ in order to manage grid congestion due to high renewable generation in these zones. The residual capacity is calculated for a whole year using RTE hourly injection data with projected installed wind capacity in $2020$. The injection data corresponds to consumption and generation data on the different nodes of the French network from $63$kV to $400$kV (around $6,000$ nodes).

\subsection{Clustering end-users}
\label{sec: emp_cluster}
\begin{figure}[htbp]
\begin{center}
\centerline{\includegraphics[scale=0.6]{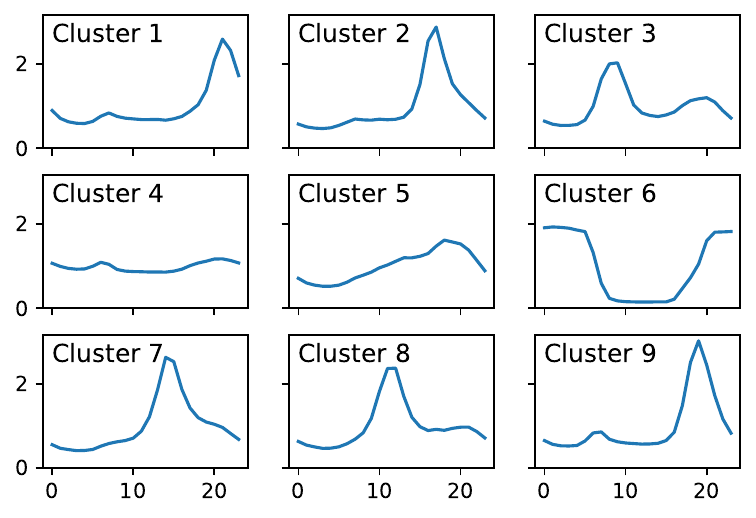}}
\caption{Daily demand profile centroids of the obtained nine clusters (normalized by the average hourly demand of the day). }
\label{fig:cluster9}
\end{center}
\end{figure}

Based on users'  demand profile data, we cluster all users into nine classes of life pattern with the following steps. First,  we cluster the daily load profiles (normalized by the average load of that day) of the customers into nine classes using k-means algorithm (The number of clusters are tuned for capturing the heterogeneity in the customer base while having enough estimation power.). The cluster centroids, as shown in Figure \ref{fig:cluster9}, are good representatives of users' life patterns especially for the time of peak energy consumption. Second, each user is  assigned to a centroid in which the user has the most daily profiles. We denote the user set of the $j$th cluster as $\mathcal{I}_j$ (i.e., $i\in\mathcal{I}_j$ if user $i$ is assigned to the $j$th class).

\subsection{Users' Optimization: Contract Choice and Virtual Battery Operation}\label{ssec: user opt empiric}
We empirically study users optimization for choosing contract and operation of the battery as in (\ref{eqn:Opt_user}). We assume that the contract cost $q^h_i$ for a user is determined according to the  large-scale battery cost in Equation \eqref{eq: battery_cost} (Residential batteries usually have fixed Energy/Power ratio, but this ratio is more flexible for large-scale battery.):
\begin{equation}
q^h_i = C^b(c^h_i, r^h_i)
\label{eq:emp_user_contrct_price}
\end{equation}
Note that this cost is affine in virtual battery size $(c^h_i, r^h_i)$. 

\begin{figure}[htbp]
\begin{center}
\centerline{\includegraphics[scale=0.7]{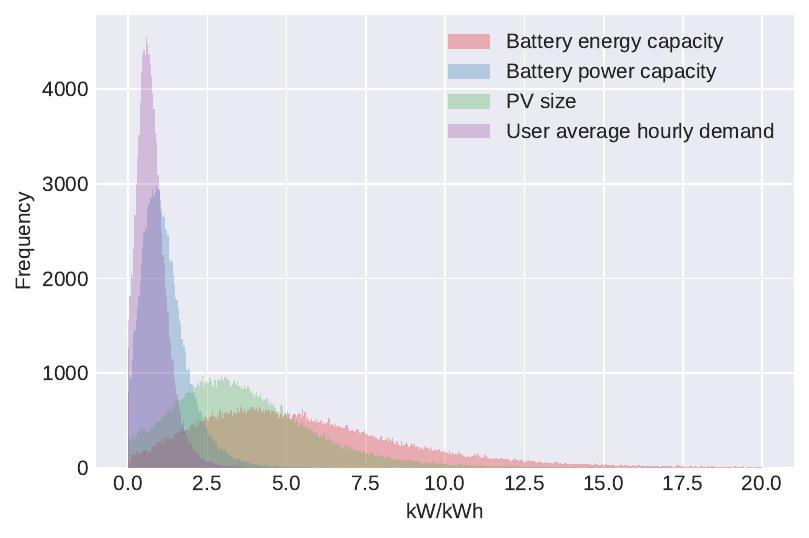}}
\caption{Distribution of energy capacity, power capacity, PV size, and average hourly demand of each user. }
\label{fig:cr_dirti}
\end{center}
\end{figure}

The user's problem defined in (\ref{eqn:Opt_user}) is then solved for all users in each type for one year. The solution to the problem includes optimal contracted energy capacity $\{c^{h*}_{i}\}$ and power capacity $\{r^{h*}_{i}\}$ for each user $i$, along with its optimal battery schedule $\{a^{h*}_{i,t}\}, t\in \{1,2,...,365\times 24\}$. The histogram of contracted energy and power capacity as well as the average hourly demand and PV size of all users are plotted in Figure \ref{fig:cr_dirti}.

\begin{figure}[htbp]
		\begin{subfigure}[b]{0.5\textwidth}
			\centerline{\includegraphics[scale=0.5]{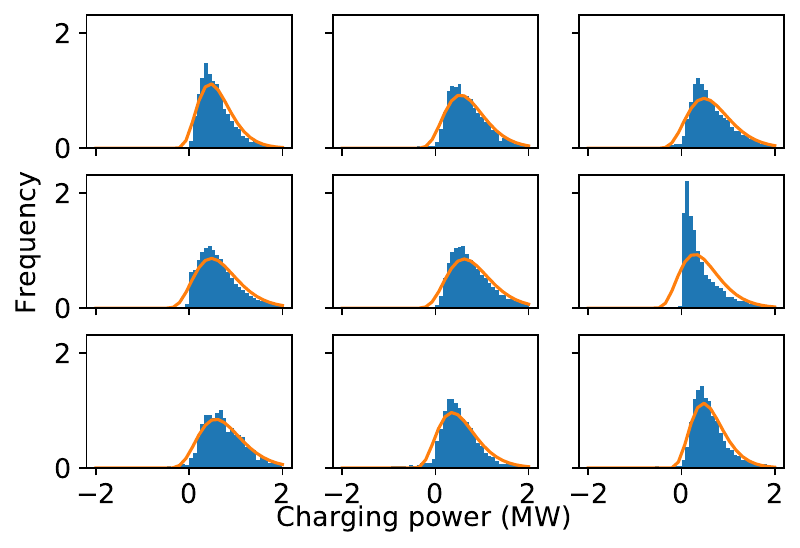}}
			\caption{10AM, a summer day}
		\end{subfigure}
		\begin{subfigure}[b]{0.5\textwidth}
			\centerline{\includegraphics[scale=0.5]{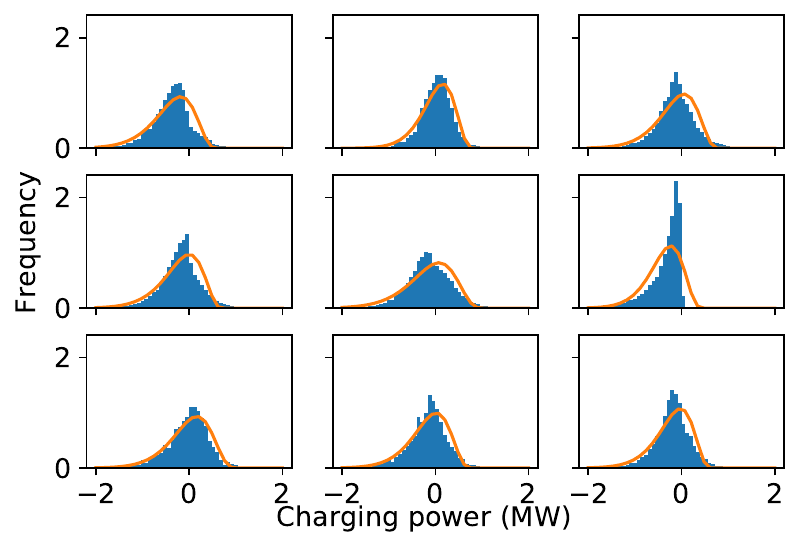}}
			\caption{10PM, a summer day}
		\end{subfigure}
		\begin{subfigure}[b]{0.5\textwidth}
			\centerline{\includegraphics[scale=0.5]{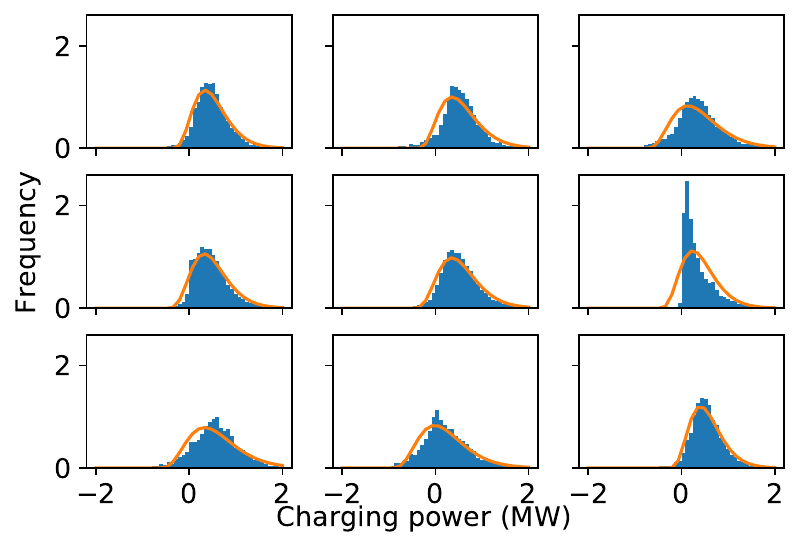}}
			\caption{10AM, a winter day}
		\end{subfigure}
		\begin{subfigure}[b]{0.5\textwidth}
			\centerline{\includegraphics[scale=0.5]{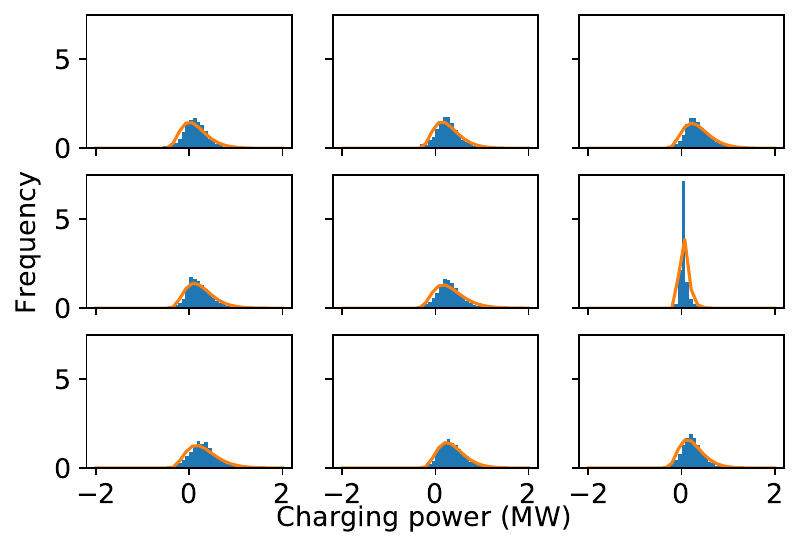}}
			\caption{10PM, a winter day}
		\end{subfigure}
		\caption{Frequency and gamma distribution fitting of charging power of the nine clusters. Two hours of two days of summer and winter are shown, respectively.}
		\label{fig:hist_battery}
	\end{figure}

 To apply  our effective capacity approximation, we fit to battery schedule of each user at each hour a gamma distribution, which is sub-exponential (By Bernstein's condition, a gamma distribution with shape parameter $\alpha$ and scale parameter $\beta$ is sub-exponential with $\nu=\sqrt{2\alpha\beta^2}$ and $b=2\beta$.). For example, a gamma distribution is fitted over $\{a^{h*}_{i,t}\},i\in \mathcal{I}_j$ for battery schedule of all users of the $j$th cluster in time $t$. For each hour we need to specify  which side, charging or discharging, of the battery schedule is the tail side for the gamma distribution. In this simulation, we choose charging as the tail side if the summed charging power is larger than the summed discharging power for a particular type of users in a particular hour of the year and vice versa. One observation we have is that the sides chosen through the described way for each type of users are the same in a particular hour. Therefore, the proposed formulation of the problem in Section \ref{sec:het_users} can be applied here. The histogram of battery schedule as well as the fitted gamma distribution for two hours in summer and winter respectively are plotted in Figure \ref{fig:hist_battery}.

\subsection{Cloud Storage: Investment and Operation}
\label{sec:cso_invest_sim}
Using the results in Section \ref{ssec: user opt empiric}, we can now study the CSO's investment and physical battery operation in (\ref{eqn:Opt_cso}). Note that (\ref{eqn:Opt_cso}) does not include congestion management, hence the constraints are deterministic. We study the multi-service of cloud storage and congestion management in Section \ref{ssec: cloud and congestion empirics}. The organization of the rest of this Section is the following. In Section \ref{ssec:with_without_ext}, we solve the CSO's problem both with and without access to external resources for balancing mismatch in battery schedule. We measure the value of the CSO business as well as our proposed operational metrics and discuss the added value of access to external resources. We compare the value of the cloud storage to alternative use cases of the battery in Section \ref{ssec:comp_other_cases}.
Then in Section \ref{ssec:sens_emp}, we do sensitivity analysis of the solution both to the number of users and to the prices in external resources (penalty prices). We also compare the results from our effective capacity approximation to those from the Monte Carlo simulation to measure the approximation gap.

Solving the proposed two-level optimization problems formulated in (\ref{eqn:Opt_user}) and (\ref{eqn:Opt_cso}) usually requires specific structures of the storage contract or relaxations of the problem sacrificing some accuracy. In this section, in order to focus on sizing problem of the cloud storage with heterogeneous users (multi-service), rather than pricing, we  suppose users face some ex-ante fixed contract plans satisfying incentive compatibility (users will not buy battery themselves) to decouple the two-level problems (\ref{eqn:Opt_user}) and (\ref{eqn:Opt_cso}). We consider each user is offered the same battery price as what the CSO faces when renting battery at scale from RTE, that is $C^b(\cdot)$ as in (\ref{eq: battery_cost}) (We assume the operation and maintenance costs of the CSO to be 0, that is, $C^b_{eq}(c^b,r^b)=C^b(c^b,r^b)$).

\subsubsection{Cloud Storage With and Without Access to External Resources} \label{ssec:with_without_ext} We study both cases of with and without access to external resources and characterize the battery investment and operation, plus the value of the cloud storage. For reporting the operation of the battery we use our proposed operational metrics in Section \ref{ssec:oper_metrics}. In this Section we use our effective capacity approximation to solve the CSO's problem.

\begin{figure}[htbp]
		\begin{subfigure}[b]{0.5\textwidth}
			\centerline{\includegraphics[scale=0.5]{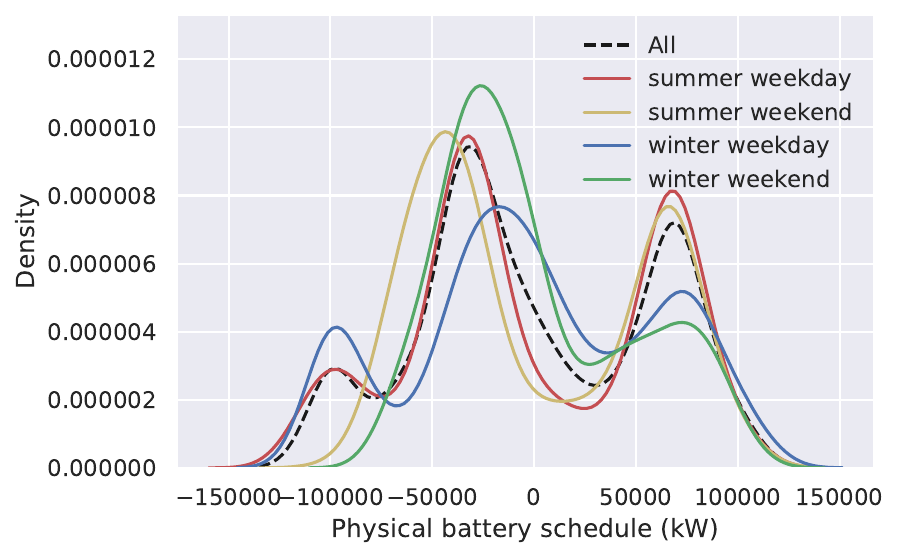}}
			\caption{Without access to external resources}
			\label{fig:aggre_1}
		\end{subfigure}
		\begin{subfigure}[b]{0.5\textwidth}
			\centerline{\includegraphics[scale=0.5]{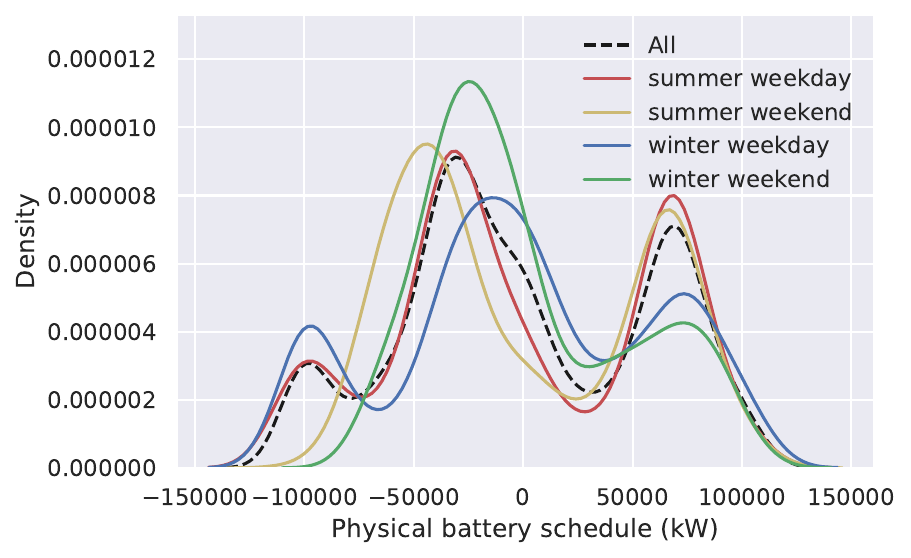}}
			\caption{With access to external resources}
			\label{fig:aggre_2}
		\end{subfigure}
		\caption{Distribution of physical battery schedule over the simulated year (dashed line) and differentiated across seasons and week days.}
		\label{fig:aggre}
	\end{figure}

\textbf{Case 1: no access of CSO to external resources} We start with a case where all 116127 users participate in the cloud storage program and the CSO is not allowed to have battery schedule mismatch, that is, facing a very high penalty price. The optimal battery energy capacity and power capacity to install in this case are equal to 657.7MWh and 121.7MW, respectively. The multiplexing gain is 3.3\%. The probability of blocking with no access to external resources is clearly zero, and this is actually the reason for low multiplexing gain. Figure \ref{fig:aggre_1} provides a more detailed view of the battery operation by plotting the distribution of the physical battery schedule. The mean is 0 with standard deviation 49.4MW for battery schedule in the whole year. The charging commands (positive values) corresponding to the summer period are centered around 60MW while for the winter these values are more smoothly spread between 0 and 70MW. This is due to a higher solar irradiance in summer and thus increased storage of surplus PV generation during the day. While all the discharge commands (negative values) present a peak between -10MW and -50MW, the distributions of weekdays show a second peak in negative values around -100MW. This secondary peak corresponds to the intensive discharge during peak hours. Note that the PG\&E tariff used for this study has flat price on weekends without peak hour pricing, which is the reason for no secondary peak on weekends. Note that in our data, there is not heterogeneity in prices because all users are under the same retail tariff. Therefore there is a high correlation in their virtual storage operation during peak periods. The expected profit of the CSO  over the simulated year in this case, which is the revenue from the users minus the equivalent battery investment cost and the blocking cost, is equal to \$ 1.5M/yr.

\textbf{Case 2: CSO with access to external resources }
We proceed with a case where all users participate in the cloud storage program and the CSO is  allowed to have battery schedule mismatch penalized at the retail ToU price. The optimal battery energy capacity and power capacity to install in this case are equal to 611.9MWh and 106.2MW, respectively. The multiplexing gain is 10\% and and the probability of blocking  is 9.6\%. 

\begin{figure}[htbp]
\begin{center}
\centerline{\includegraphics[scale=0.7]{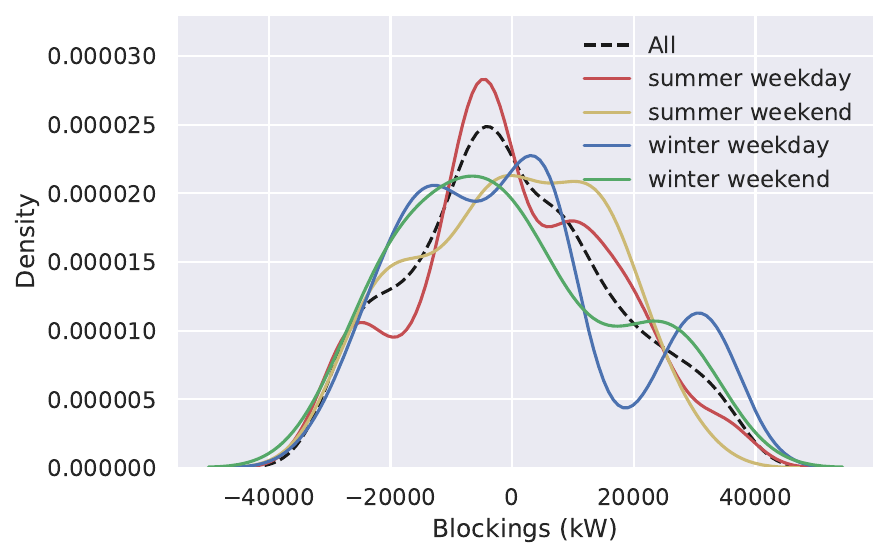}}
\caption{Distribution of blockings $\sum_i a_{i,t}^{h*}-a_t^b$ over the simulated year (dashed line) and differentiated  seasons and day types. }
\label{fig:blockings}
\end{center}
\end{figure}

{Figure \ref{fig:aggre_2} and Figure \ref{fig:blockings} provide more detailed view of the operation of the battery by plotting distribution of the physical battery schedule and the blocking values. Figure \ref{fig:aggre_2} shows that  charging and discharging commands with access to external resources are centered around almost the same centroids as those in Figure \ref{fig:aggre_1} though the densities may vary.} Figure \ref{fig:blockings} shows a mean blocking value equal to 0 and a standard deviation equal to 15.6MW.  The positive values correspond to the excess the CSO cannot charge into the battery (positive blockings are not penalized), while the negative values correspond to the missing energy the CSO needs to get from external resources. While blockings over the year are centered around -2.5MW, the distributions of blockings of different seasons and day types vary. In general, the distributions of blockings in winter appear to  have more peaks than in summer. One reason for this is that solar panels have more fluctuating output in winter than in summer due to the weather and that solar generation has a high correlation with charging patterns. The expected profit of the CSO  over the simulated year in this case is equal to \$ 2.8M/yr.

We now compare Case 1 and Case 2. In Case 1 the CSO invests on more physical battery because he can not rely on the external resources. Both Case 1 and Case 2 show that there are noticeable multiplexing gain obtained from heterogeneous battery usage. Access to external resources allows an additional 6.3\% multiplexing gain. {Comparing Figure \ref{fig:aggre_1} and Figure \ref{fig:aggre_2} for more detailed comparison of the battery operation shows that with access to external resources the densities of discharging peaks reduce in summer while there is almost no changed in winter. This aligns with the previous finding that discharging tends to be larger in summer than in winter due to solar irradiation difference and the intuition that blocking tends to happen when discharging is higher}. Finally, access to external resources provides an additional yearly profit of \$1.3M/yr, about $85\%$ improvement to lack of external resources. Also the profit per unit of investment is $\$12k/MW$ for Case 1 and $\$26k/MW$ in Case 2, $116\%$ improvement.

\subsubsection{Comparison to Alternative Services} 
\label{ssec:comp_other_cases}

We compare CSO operation to other battery services studied in the literature. PG\&E performed a study to assess the economic performance of two of its large scale batteries from their participation in the California Independent System Operator (CAISO) markets, also known as the EPIC Project \citep{fribush2016epic}. This real-world experiment showed that revenues from energy arbitrage on the wholesale market were too low to compensate for the inherent round trip efficiency of the batteries, with a yearly profit of around \$8.5k/MW. The revenues from frequency regulation represented the highest value, with the yearly profit ranging between \$24k and \$84k/MW. These values are  transformation of Case 2 will lead to a yearly profit of around \$26k/MW, which is in the same order as that from frequency regulation service. This is while the frequency regulation is a high priority service that can not be mixed with congestion management.

\subsubsection{Sensitivity Analysis}
\label{ssec:sens_emp}
We analyze the sensitivity of the battery operation and planning to cost of external resources in covering blocking occasions, as well as to the number of users. We also assess the gap between our effective capacity approximation and Monte Carlo solution to the optimization problems. To this end, we solve the two-stage optimization problem by changing external costs from very small to very large values and for two cases of 10k users and 100k users (almost all of the users) respectively (We assume that the proportion of each type of users is the same as that in our whole dataset, but this assumption can be easily relaxed by specifying an arbitrary number for each type of users.). For each setup we solve both with our effective capacity approximation and with Monte Carlo simulation.

\begin{figure}[htbp]
		\begin{subfigure}[b]{0.5\textwidth}
			\centerline{\includegraphics[scale=0.5]{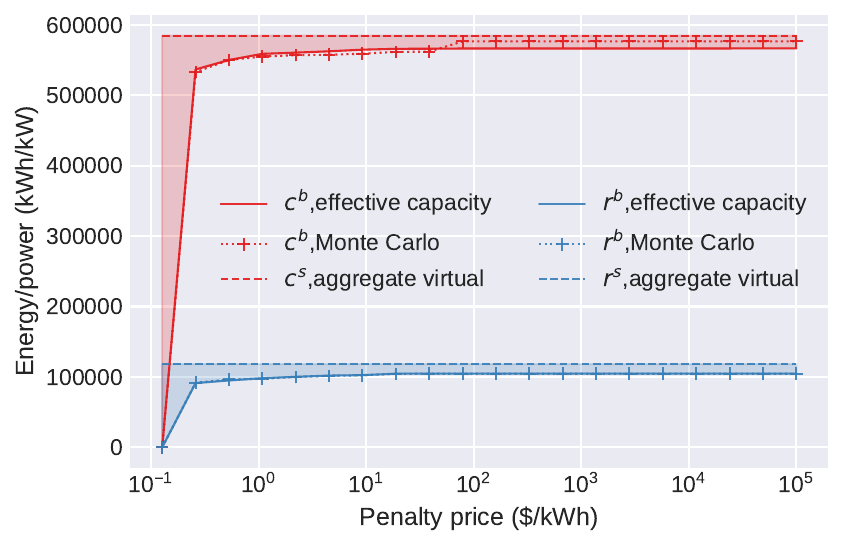}}
			\caption{Battery energy and power capacity}
			\label{fig:100k_1}
		\end{subfigure}
		\begin{subfigure}[b]{0.5\textwidth}
			\centerline{\includegraphics[scale=0.5]{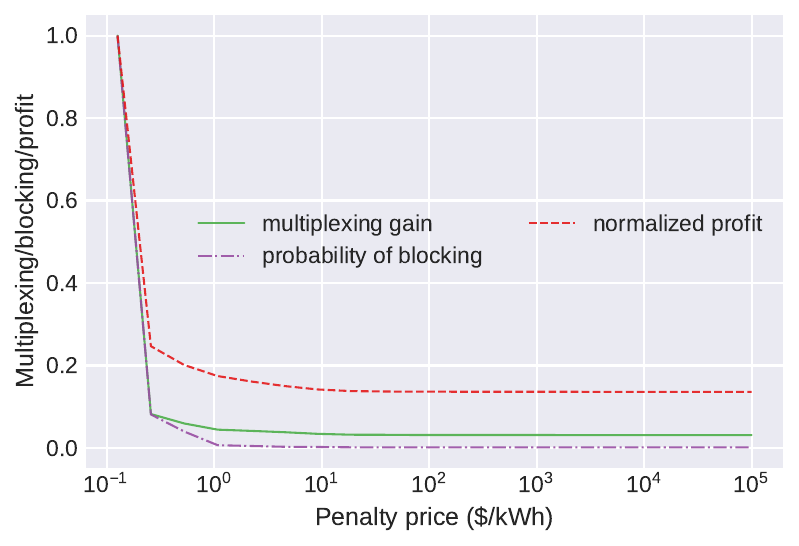}}
			\caption{Multiplexing gain, blocking, normalized profit}
			\label{fig:100k_2}
		\end{subfigure}
		\caption{Energy capacity, power capacity, multiplexing gain, probability of blocking, and normalized CSO profit for 100k users. The CSO yearly profit is normalized by zero external cost profit.}
		\label{fig:100k}
	\end{figure}

Figure \ref{fig:100k} shows the results for 100k users under varying costs of external resources. In Figure \ref{fig:100k_1}, we present the optimal battery investment for both energy and power capacity using both Monte Carlo solution and effective capacity approximation. First observation is that the results show a very low approximation gap. Second, the investment on power and energy increases with higher penalty costs because the marginal value of investment is binded with the penalty cost and the multiplexing gain; with higher costs the first constraint is uplifted. Furthermore, this figure shows the energy and power capacity of the aggregate virtual battery provided to the end-users. Note that they are both constant and indifferent to the cost of external resources since we have fixed the contract menu in these numerical studies. At the limit the investment becomes close to the aggregate virtual battery, and the difference is due to multiplexing gain. Figure \ref{fig:100k_2} shows how  multiplexing gain, probability of blocking, and CSO's profit change with varying penalty prices. As the penalty price increases, all three metrics decrease. At the limit, when the external resource are too expensive to use, we observe no blocking event, low multiplexing gain (merely from statistical multiplexing), and a profit only from statistical multiplexing. The findings here are all aligned with our findings in Case 1 of Section \ref{ssec:with_without_ext} where the probability of blocking was 0. Note however that in that case we assumed the external resource prices are the time of use prices with peak and off-peak, rather than the constant external prices here.

\begin{figure}[htbp]
		\begin{subfigure}[b]{0.5\textwidth}
			\centerline{\includegraphics[scale=0.5]{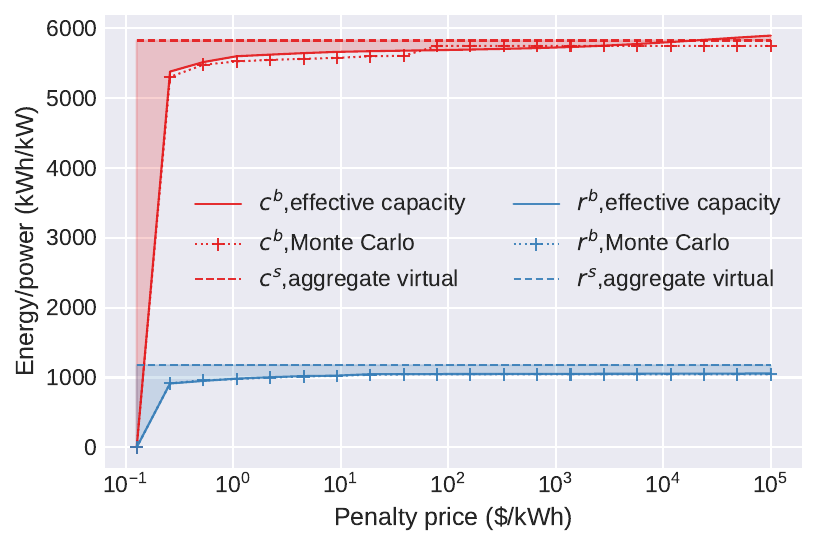}}
			\caption{Battery energy and power capacity}
			\label{fig:1k_1}
		\end{subfigure}
		\begin{subfigure}[b]{0.5\textwidth}
			\centerline{\includegraphics[scale=0.5]{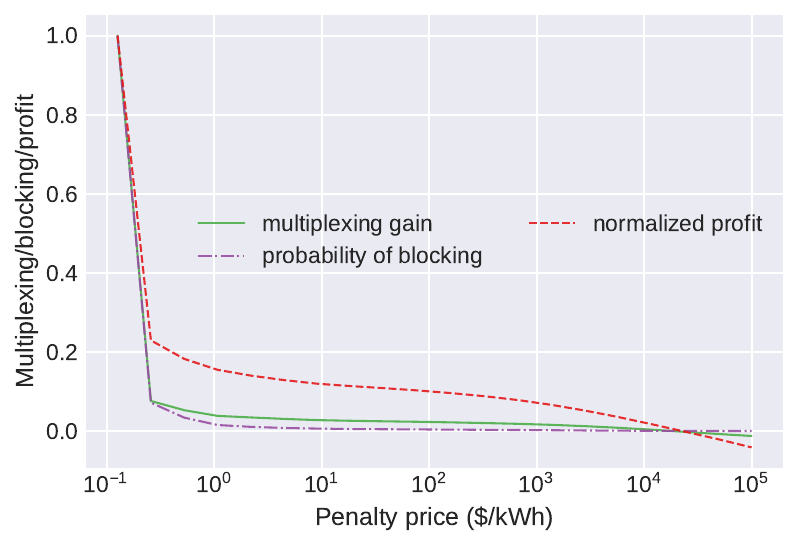}}
			\caption{Multiplexing gain, blocking,  normalized profit}
			\label{fig:1k_2}
		\end{subfigure}
		\caption{Energy capacity, power capacity, multiplexing gain, probability of blocking, and normalized CSO profit for 1k users. The CSO yearly profit is normalized by the zero external cost profit.}
		\label{fig:1k}
	\end{figure}

\begin{figure}[htbp]
\begin{center}
\centerline{\includegraphics[scale=0.5]{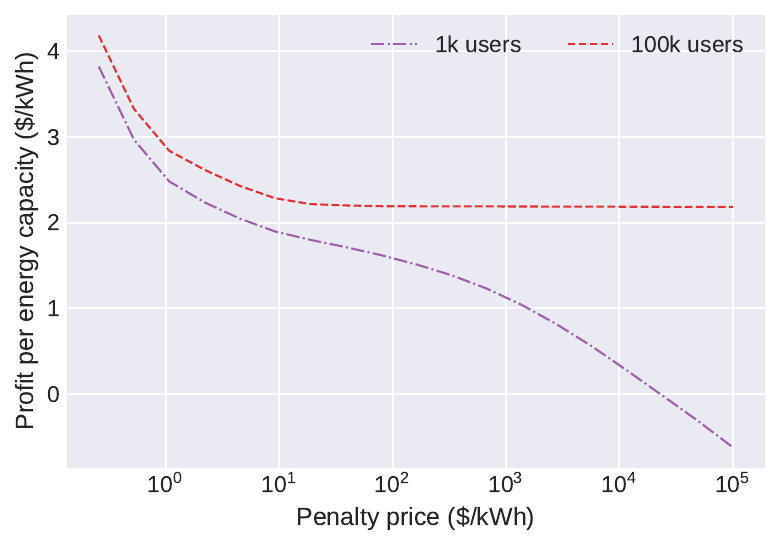}}
\caption{Profit per energy capacity for 1k and 100k users as a function of penalty cost of external resources.}
\label{fig:profit_mwh}
\end{center}
\end{figure}

Similarly, Figure \ref{fig:1k} shows the results for 1k users under varying costs of external resources. While most observations are the same as in the case of 100k users, comparing Figure \ref{fig:1k} with \ref{fig:100k} shows that as expected the multiplexing gain is lower with lower number of users. The profit with 1k user is also more sensitive to cost of external resources. Figure \ref{fig:profit_mwh} shows the CSO's profit per unit of battery invested for both 1k and 100k users. It is always larger for 100k users vs 1k and for both it decreases with larger penalty prices. Due to higher multiplexing gain the profit for 100k converges to some positive values at high penalty costs (with zero blocking probability), but for 1k with low multiplexing gain the profit with zero blocking probability becomes negative. 

\subsection{Cloud Storage with Congestion Management}\label{ssec: cloud and congestion empirics}
In this Section, we empirically assess the effect of congestion management on the CSO's investment problem. To do this, we extend the treatment in Section \ref{sec:cso_invest_sim} by replacing deterministic capacity constraints of (\ref{eq:power_cap_cons}) and (\ref{eq:energy_cap_cons})   with stochastic constraints of (\ref{eq: residual power}) and (\ref{eq: residual energy}). Our solution approach, as discussed at the end of Section \ref{ssec: problems}, is using chance constraints of (\ref{eq: chance_cons}).
The results show that congestion management results a reduced investment (leasing) level and expected profit for CSO. Nevertheless, if the battery owner (e.g. RTE) provides discount in cost of access to the stochastic battery $C^b(c^b, r^b)$ by 1.2\% compared to a deterministic battery, then CSO's profit will resume to the results in the previous Section.

The problem and setup in this section is similar to the one studied in Case 2 of Section \ref{ssec:with_without_ext} of course with chance constraints of (\ref{eq: chance_cons}). We set $\eta=0.9$. Also, we form the distribution of $r_t^{b,l}$, $r_t^{b,h}$, and $c_t^b$ by using RTE data for hourly grid congestion events over a year provided in \cite{straub2018congestion} and scale accordingly. In this realization, the battery capacity is fully available 87\% of the time across the simulated year, and the full capacity of the battery is used for congestion management only 5\% of the time. We form an empirical hourly distribution per season by averaging over all congestion events over that hour during a season. The non-congested hours are assumed to be always fully available.

\begin{figure}[htbp]
\begin{center}
\centerline{\includegraphics[scale=0.7]{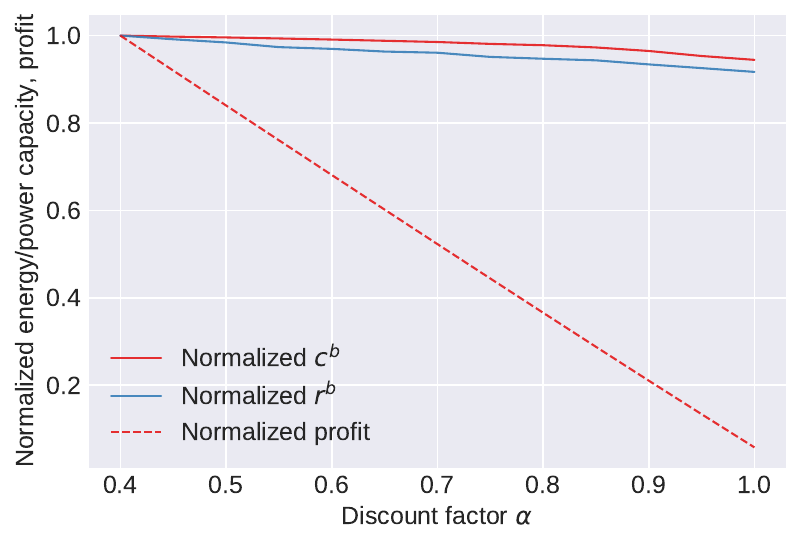}}
\caption{Energy capacity, power capacity, and CSO's profit normalized to their maximum values, under congestion management with varying battery cost and $\eta=0.9$.}
\label{fig:congest_invest}
\end{center}
\end{figure}

Figure \ref{fig:congest_invest} shows energy capacity, power capacity, and CSO's profit normalized to their maximum values (638.4MWh, 115.9MW, \$ 36.5M/year respectively), under varying stochastic battery leasing cost $\tilde{C}^{b}$ to the CSO with
\begin{equation}
\tilde{C}^{b}(r^b,c^b) = \alpha C^b(r^b,c^b)
\end{equation}
where $\alpha$ is the discount factor and $C^b(r^b, c^b)$ is from (\ref{eq: battery_cost}). The first observation is that energy capacity, power capacity, and profit all decrease with increasing battery leasing cost. Second, with $\alpha=1$, i.e. the cost of stochastic battery is equal to that of the deterministic one,  energy capacity, power capacity, and profit are $603.0$MWh, $106.2$M, and \$ $2.1$M/year respectively. Compared to Case 2 of Section \ref{ssec:with_without_ext} where congestion management is not considered, the energy capacity invested drops by 1.5\%, the power capacity remains almost unchanged, and the profit drops by 25\% because the availability of battery capacity is constrained by the high priority congestion management.  
However, this profit of \$ 2.1M/year does not account for the additional value added from congestion management service which is required for assessing the value of the battery in multi services. The value of battery grid service is hard to assess and is sensitive to its location. \cite{rte2017rei} provides values between \$10.60k/MWh$\cdot$yr to \$68.87k/MWh$\cdot$yr, which results in a profit of \$$2.7$M/yr to \$$4.1$M/yr from congestion management from CSO's level of investment under $\alpha=1$. Thus, the value from congestion management is higher than the profit drop, making this multi-service battery economically interesting.  
Third, with only a slight discount $\alpha=0.988$ in the battery leasing cost, the profit of the CSO with congestion management can be resumed to that without congestion management, that is, \$ 2.8M/year.

\section{Conclusion and future work} \label{sec:conclusion}
We studied cloud electricity storage for multi-service battery operation with high priority grid services. We proposed a multi-service multi-user architecture for the problem and operation metrics of multiplexing gain and probability of blocking. To address computational challenges, we proposed effective capacity for approximating CSO's blocking penalties and this way forming a convex approximation to CSO's profit. 

We conducted empirical analysis using the case of RTE, where the battery is primarily for congestion management and the residual battery is rented to a third-party that uses it for providing cloud storage service to end-users. Our empirical study shows CSO of the residual battery is an economically interesting business if there is enough diversity in the users profiles to increase the statistical multiplexing, and if the cost of external resources are affordable. We provided sensitivity analysis to these two factors, as well as the RTE leasing price of the battery to the CSO. Our empirical study also shows our proposed effective capacity is a close approximation.

This work can be extended in several directions. Our empirical study can be replicated with France end-user data (we did not have access to such data). Another direction is modeling the retailers' price response to CSO's market entrance. Our framework can also be adapted for other multi-user multi-service battery operations.

\ACKNOWLEDGMENT{%
}

%
%
%


\bibliographystyle{informs2014} 
\bibliography{sample.bib}


\end{document}